\documentclass[pra,twocolumn,showpacs,a4paper]{revtex4}
\usepackage{amssymb}
\usepackage{epsfig}
\usepackage{amsmath}
\usepackage{graphicx}
\usepackage{bm}

\begin{document}

\title{Robust entanglement of a micromechanical resonator with output optical fields}
\author{C. Genes, A. Mari, P. Tombesi, and D. Vitali}
\affiliation{Dipartimento di Fisica, Universit\`{a} di Camerino, I-62032 Camerino (MC), Italy}

\begin{abstract}
We perform an analysis of the optomechanical entanglement between the experimentally detectable output field of an optical cavity and a
vibrating cavity end-mirror. We show that by a proper choice of the readout (mainly by a proper choice of detection bandwidth) one can not only
detect the already predicted intracavity entanglement but also optimize and increase it. This entanglement is explained as being generated by a
scattering process owing to which strong quantum correlations between the mirror and the optical Stokes sideband are created. All-optical
entanglement between scattered sidebands is also predicted and it is shown that the mechanical resonator and the two sideband modes form a fully
tripartite-entangled system capable of providing practicable and robust solutions for continuous variable quantum communication protocols.
\end{abstract}


\maketitle

\section{Introduction}

Mechanical resonators at the micro- and nano-meter scale are now widely employed in the high-sensitive detection of mass and forces
\cite{blencowe,roukpt,optexpr}. The recent improvements in the nanofabrication techniques suggest that in the near future these devices will
reach the regime in which their sensitivity will be limited by the ultimate quantum limits set by the Heisenberg principle, as first suggested
in the context of the detection of gravitational waves by the pioneering work of Braginsky and coworkers \cite{bragbook}.

The experimental demonstration of genuine quantum states of macroscopic mechanical resonators with a mass in the nanogram-milligram range will
represent an important step not only for the high-sensitive detection of displacements and forces, but also for the foundations of physics. It
would represent, in fact, a remarkable signature of the quantum behavior of a macroscopic object, allowing to shed further light onto the
quantum-classical boundary \cite{found}. Significant experimental
\cite{cohadon99,schwab,karrai,naik,arcizet06,gigan06,arcizet06b,bouwm,vahalacool,mavalvala,rugar,wineland,markusepl,sidebcooling,harris,lehnert}
and theoretical \cite{Mancini98,brag,courty,quiescence02,imazoller,tianzoller,marquardt,wilson-rae,genes07,dantan07} efforts are currently
devoted to cooling such microresonators to their quantum ground state.

However, the generation of other examples of quantum states of a micro-mechanical resonator has been also considered recently. The most relevant
examples are given by squeezed and entangled states. Squeezed states of nano-mechanical resonators \cite{blencowe-wyb} are potentially useful
for surpassing the standard quantum limit for position and force detection \cite{bragbook}, and could be generated in different ways, using
either the coupling with a qubit \cite{squee1}, or measurement and feedback schemes \cite{quiescence02,squee2}. Entanglement is instead the
characteristic element of quantum theory, because it is responsible for correlations between observables that cannot be understood on the basis
of local realistic theories \cite{Bell64}. For this reason, there has been an increasing interest in establishing the conditions under which
entanglement between macroscopic objects can arise. Relevant experimental demonstration in this directions are given by the entanglement between
collective spins of atomic ensembles~\cite{Polzik}, and between Josephson-junction qubits~\cite{Berkley}. Then, starting from the proposal of
Ref.~\cite{PRL02} in which two mirrors of a ring cavity are entangled by the radiation pressure of the cavity mode, many proposals involved
nano- and micro-mechanical resonators, eventually entangled with other systems. One could entangle a nanomechanical oscillator with a
Cooper-pair box \cite{Armour03}, while Ref.~\cite{eisert} studied how to entangle an array of nanomechanical oscillators. Further proposals
suggested to entangle two charge qubits \cite{zou1} or two Josephson junctions \cite{cleland1} via nanomechanical resonators, or to entangle two
nanomechanical resonators via trapped ions \cite{tian1}, Cooper pair boxes \cite{tian2}, or dc-SQUIDS \cite{nori}. More recently, schemes for
entangling a superconducting coplanar waveguide field with a nanomechanical resonator, either via a Cooper pair box within the waveguide
\cite{ringsmuth}, or via direct capacitive coupling \cite{Vitali07}, have been proposed.

After Ref.~\cite{PRL02}, other optomechanical systems have been proposed for entangling optical and/or mechanical modes by means of the
radiation pressure interaction. Ref.~\cite{Peng03} considered two mirrors of two different cavities illuminated with entangled light beams,
while Refs.~\cite{pinard-epl,meystre,paternostro,wipf} considered different examples of double-cavity systems in which entanglement either
between different mechanical modes, or between a cavity mode and a vibrational mode of a cavity mirror have been studied.
Refs.~\cite{prl07,jopa} considered the simplest scheme capable of generating stationary optomechanical entanglement, i.e., a single Fabry-Perot
cavity either with one \cite{prl07}, or both \cite{jopa}, movable mirrors.

Here we shall reconsider the Fabry-Perot model of Ref.~\cite{prl07}, which is remarkable for its simplicity and robustness against temperature
of the resulting entanglement, and extend its study in various directions. In fact, entangled optomechanical systems could be profitably used
for the realization of quantum communication networks, in which the mechanical modes play the role of local nodes where quantum information can
be stored and retrieved, and optical modes carry this information between the nodes. Refs.~\cite{prltelep,jmo,Pir06} proposed a scheme of this
kind, based on free-space light modes scattered by a single reflecting mirror, which could allow the implementation of continuous variable (CV)
quantum teleportation \cite{prltelep}, quantum telecloning \cite{jmo}, and entanglement swapping \cite{Pir06}. Therefore, any quantum
communication application involves \emph{traveling output} modes rather than intracavity ones, and it is important to study how the
optomechanical entanglement generated within the cavity is transferred to the output field. Furthermore, by considering the output field, one
can adopt a multiplexing approach because, by means of spectral filters, one can always select many different traveling output modes originating
from a single intracavity mode (see Fig.~1). One can therefore manipulate a multipartite system, eventually possessing multipartite
entanglement. We shall develop a general theory showing how the entanglement between the mechanical resonator and optical output modes can be
properly defined and calculated.

We shall see that, together with its output field, the single Fabry-Perot cavity system of Ref.~\cite{prl07} represents the ``cavity version''
of the free-space scheme of Refs.~\cite{prltelep,jmo}. In fact, as it happens in this latter scheme, all the relevant dynamics induced by
radiation pressure interaction is carried by the two output modes corresponding to the first Stokes and anti-Stokes sidebands of the driving
laser. In particular, the optomechanical entanglement with the intracavity mode is optimally transferred to the output Stokes sideband mode,
which is however robustly entangled also with the anti-Stokes output mode. We shall see that the present Fabry-Perot cavity system is preferable
with respect to the free space model of Refs.~\cite{prltelep,jmo}, because entanglement is achievable in a much more accessible experimental
parameter region.

The outline of the paper is as follows. Sec.~II gives a general description of the dynamics by means of the Quantum Langevin Equations (QLE),
Sec.~III analyzes in detail the entanglement between the mechanical mode and the intracavity mode, while in Sec.~IV we describe a general theory
on how a number of independent optical modes can be selected and defined, and their entanglement properties calculated. Sec. V is for concluding
remarks.

\section{System dynamics}

We consider a driven optical cavity coupled by radiation pressure to a micromechanical oscillator. The typical experimental configuration is a
Fabry-Perot cavity with one mirror much lighter than the other (see e.g. \cite{karrai,gigan06,arcizet06,arcizet06b,bouwm,harris}), but our
treatment applies to other configurations such as the silica toroidal microcavity of Refs.~\cite{vahalacool,sidebcooling,vahala1}. Radiation
pressure couples each cavity mode with many vibrational normal modes of the movable mirror. However, by choosing the detection bandwidth so that
only an isolated mechanical resonance significantly contributes to the detected signal, one can restrict to a single mechanical oscillator,
since inter-mode coupling due to mechanical nonlinearities are typically negligible (see also \cite{duemodi} for a more general treatment). The
Hamiltonian of the system reads \cite{GIOV01}
\begin{eqnarray}
&& H=\hbar\omega_{c}a^{\dagger}a+\frac{1}{2}\hbar\omega_{m}(p^{2}+q^{2})-
\hbar G_{0}a^{\dagger}a q  \\
&& +i\hbar E(a^{\dagger}e^{-i\omega_{0}t}-ae^{i\omega_{0}t}).  \label{ham0}
\end{eqnarray}
The first term describes the energy of the cavity mode, with lowering operator $a$ ($[a,a^{\dag}]=1$), cavity frequency $\omega_c$ and decay
rate $ \kappa$. The second term gives the energy of the mechanical mode, modeled as harmonic oscillator at frequency $\omega_m$ and described by
dimensionless position and momentum operators $q$ and $p$ ($ [q,p]=i$). The third term is the radiation-pressure coupling of rate $
G_0=(\omega_c/L)\sqrt{\hbar/m \omega_m}$, where $m$ is the effective mass of the mechanical mode \cite{Pinard}, and $L$ is an effective length
that depends upon the cavity geometry: it coincides with the cavity length in the Fabry-Perot case, and with the toroid radius in the case of
Refs.~\cite{vahalacool,vahala1}. The last term describes the input driving by a laser with frequency $\omega_0$, where $E$ is related to the
input laser power $P$ by $|E|=\sqrt{2P \kappa/\hbar \omega_0}$. One can adopt the single cavity mode description of Eq.~(\ref{ham0}) as long as
one drives only one cavity mode and the mechanical frequency $\omega_m$ is much smaller than the cavity free spectral range $FSR \sim c/L$. In
this case, scattering of photons from the driven mode into other cavity modes is negligible \cite{law}.

\begin{figure}[tb]\label{fig1}
\centerline{\includegraphics[width=0.45\textwidth]{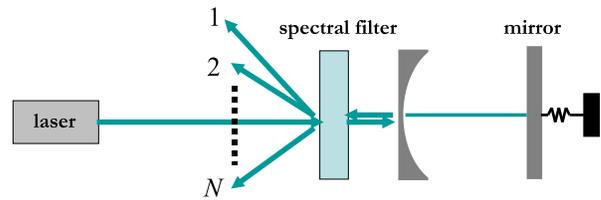}} \caption{Scheme of the cavity, which is driven by a laser and has a vibrating
mirror. With appropriate filters one can select $N$ independent modes from the cavity output field.}
\end{figure}

The dynamic is also determined by the fluctuation-dissipation processes affecting both the optical and the mechanical mode. They can be taken
into account in a fully consistent way \cite{GIOV01} by considering the following set of nonlinear QLE, written in the interaction picture with
respect to $ \hbar \omega_0 a^{\dag}a$
\begin{subequations}
\label{nonlinlang}
\begin{eqnarray}
\dot{q}&=&\omega_m p, \\
\dot{p}&=&-\omega_m q - \gamma_m p + G_0 a^{\dag}a + \xi, \\
\dot{a}&=&-(\kappa+i\Delta_0)a +i G_0 a q +E+\sqrt{2\kappa} a^{in},
\end{eqnarray}
where $\Delta_0=\omega_c-\omega_0$. The mechanical mode is affected by a viscous force with damping rate $\gamma_m$ and by a Brownian stochastic
force with zero mean value $\xi$, that obeys the correlation function \cite{Landau,GIOV01}
\end{subequations}
\begin{equation}  \label{browncorre}
\left \langle \xi(t) \xi(t^{\prime})\right \rangle = \frac{\gamma_m}{\omega_m%
} \int \frac{d\omega}{2\pi} e^{-i\omega(t-t^{\prime})} \omega \left[%
\coth\left(\frac{\hbar \omega}{2k_BT}\right)+1\right],
\end{equation}
where $k_B$ is the Boltzmann constant and $T$ is the temperature of the reservoir of the micromechanical oscillator. The Brownian noise $\xi(t)$
is a Gaussian quantum stochastic process and its non-Markovian nature (neither its correlation function nor its commutator are proportional to a
Dirac delta) guarantees that the QLE of Eqs.~(\ref{nonlinlang}) preserve the correct commutation relations between operators during the time
evolution \cite{GIOV01}. The cavity mode amplitude instead decays at the rate $\kappa$ and is affected by the vacuum radiation input noise
$a^{in}(t)$, whose correlation functions are given by
\begin{equation}
\langle a^{in}(t)a^{in,\dag}(t^{\prime})\rangle =\left[N(\omega_c)+1\right]%
\delta (t-t^{\prime}),  \label{input1}
\end{equation}
and
\begin{equation} \langle a^{in,\dag}(t)a^{in}(t^{\prime})\rangle =N(\omega_c)\delta (t-t^{\prime}), \label{input2}
\end{equation}
where $N(\omega_c)=\left(\exp\{\hbar \omega_c/k_BT\}-1\right)^{-1}$ is the equilibrium mean thermal photon number. At optical frequencies $\hbar
\omega_c/k_BT \gg 1$ and therefore $N(\omega_c)\simeq 0$, so that only the correlation function of Eq.~(\ref{input1}) is relevant. We shall
neglect here technical noise sources, such as the amplitude and phase fluctuations of the driving laser. They can hinder the achievement of
genuine quantum effects (see e.g. \cite{sidebcooling}), but they could be easily accounted for by introducing fluctuations of the modulus and of
the phase of the driving parameter $E$ of Eq.~(\ref{ham0}) \cite{marin}.

\subsection{Linearization around the classical steady state and stability analysis}

As shown in \cite{prl07}, significant optomechanical entanglement is achieved when radiation pressure coupling is strong, which is realized when
the intracavity field is very intense, i.e., for high-finesse cavities and enough driving power. In this limit (and if the system is stable) the
system is characterized by a semiclassical steady state with the cavity mode in a coherent state with amplitude $ \alpha_s = E/(\kappa+i
\Delta)$, and the micromechanical mirror displaced by $q_s = G_0 |\alpha_s|^2/\omega_m$ (see Refs.~\cite{prl07,genes07,manc-tomb} for details).
The expression giving the intracavity amplitude $\alpha_s$ is actually an implicit nonlinear equation for $\alpha_s$ because
\begin{equation}
\Delta = \Delta_0- \frac{G_0^2 |\alpha_s|^2}{\omega_m}.
\end{equation}
is the effective cavity detuning including the effect of the stationary radiation pressure. As shown in Refs.~\cite{prl07,genes07}, when
$|\alpha_s| \gg 1$ the quantum dynamics of the fluctuations around the steady state is well described by linearizing the nonlinear QLE of
Eqs.~(\ref{nonlinlang}). Defining the cavity field fluctuation quadratures $\delta X\equiv(\delta a+\delta a^{\dag})/\sqrt{2}$ and $\delta
Y\equiv(\delta a-\delta a^{\dag})/i\sqrt{2}$, and the corresponding Hermitian
input noise operators $X^{in}\equiv(a^{in}+a^{in,\dag})/\sqrt{2}$ and $%
Y^{in}\equiv(a^{in}-a^{in,\dag})/i\sqrt{2}$, the linearized QLE can be written in the following compact matrix form \cite{prl07}
\begin{equation}\label{compaeq}
\dot{u}(t)=A u(t)+n(t),
\end{equation}
where $u^{T}(t) =(\delta q(t), \delta p(t),\delta X(t), \delta Y(t))^T$ ($^{T}$ denotes the transposition) is the vector of CV fluctuation
operators , $n^{T}(t) =(0, \xi(t),\sqrt{2\kappa}X^{in}(t), \sqrt{2\kappa}Y^{in}(t))^T$ the corresponding vector of noises and $A$ the matrix
\begin{equation}\label{dynmat}
  A=\left(\begin{array}{cccc}
    0 & \omega_m & 0 & 0 \\
     -\omega_m & -\gamma_m & G & 0 \\
    0 & 0 & -\kappa & \Delta \\
    G & 0 & -\Delta & -\kappa
  \end{array}\right),
\end{equation}
where
\begin{equation} G=G_0 \alpha_s\sqrt{2}=\frac{2\omega_c}{L}\sqrt{\frac{P \kappa}{m \omega_m \omega_0 \left(\kappa^2+\Delta^2\right)}},
\label{optoc}
\end{equation}
is the \emph{effective} optomechanical coupling (we have chosen the phase reference so that $\alpha_s$ is real and positive). When $\alpha_s \gg
1$, one has $G \gg G_0$, and therefore the generation of significant optomechanical entanglement is facilitated in this linearized regime.

The formal solution of Eq.~(\ref{compaeq}) is $ u(t)=M(t) u(0)+\int_0^t ds M(s) n(t-s)$, where $M(t)=\exp\{A t\}$. The system is stable and
reaches its steady state for $t \to \infty$ when all the eigenvalues of $A$ have negative real parts so that $M(\infty)=0$. The stability
conditions can be derived by applying the Routh-Hurwitz criterion \cite{grad}, yielding the following two nontrivial conditions on the system
parameters,
\begin{subequations}
\label{stab}
\begin{eqnarray}
&& s_1=2\gamma_m\kappa\left\{ \left[ \kappa^{2}+\left( \omega_m-\Delta\right) ^{2}\right] \left[ \kappa^{2}+\left(
\omega_m+\Delta\right) ^{2}\right] \right.  \notag \\
&&\left.+\gamma_m\left[ \left( \gamma_m+2\kappa\right) \left( \kappa^{2}+\Delta ^{2}\right) +2\kappa\omega_{m}^{2}\right] \right\}  \notag
\\
&&+\Delta\omega_{m} G^{2}\left( \gamma_m+2\kappa\right) ^{2}>0, \\
&&s_2=\omega_{m}\left( \kappa^{2}+\Delta^{2}\right) -G^{2}\Delta>0.
\end{eqnarray}
\end{subequations}
which will be considered to be satisfied from now on. Notice that when $\Delta > 0$ (laser red-detuned with respect to the cavity) the first
condition is always satisfied and only $s_2$ is relevant, while when $\Delta < 0$ (blue-detuned laser), the second condition is always satisfied
and only $s_1$ matters.

\subsection{Correlation matrix of the quantum fluctuations of the system}

The steady state of the bipartite quantum system formed by the vibrational mode of interest and the fluctuations of the intracavity mode can be
fully characterized. In fact, the quantum noises $\xi$ and $a^{in}$ are zero-mean quantum Gaussian noises and the dynamics is linearized, and as
a consequence, the steady state of the system is a zero-mean bipartite Gaussian state, fully characterized by its $4 \times 4 $ correlation
matrix (CM) $ V_{ij}=\left(\langle u_i(\infty)u_j(\infty)+ u_j(\infty)u_i(\infty)\rangle\right)/2$. Starting from Eq.~(\ref{compaeq}), this
steady state CM can be determined in two equivalent ways. Using the Fourier transforms $\tilde{u}_i(\omega)$ of $u_i(t)$, one has
\begin{equation}
V_{ij}(t)=\int \int\frac{d\omega d\omega'}{4\pi}e^{-it(\omega+\omega')}\left\langle \tilde{u}_i(\omega) \tilde{u}_j(\omega')+
\tilde{u}_j(\omega')\tilde{u}_i(\omega)\right\rangle. \label{defV}
\end{equation}
Then, by Fourier transforming Eq.~(\ref{compaeq}) and the correlation functions of the noises, Eqs.~(\ref{browncorre}) and (\ref{input1}), one
gets
\begin{eqnarray}\label{lemma}
&&\frac{\left\langle \tilde{u}_i(\omega) \tilde{u}_j(\omega')+ \tilde{u}_j(\omega')\tilde{u}_i(\omega)\right\rangle}{2} \\
&&= \left[\tilde{M}(\omega)D(\omega)\tilde{M}(\omega')^{T}\right]_{ij}\delta(\omega+\omega'),
\end{eqnarray}
where we have defined the $4 \times 4 $ matrices
\begin{equation} \tilde{M}(\omega)=\left(i\omega+A\right)^{-1} \label{invea}
\end{equation}
and
\begin{equation}\label{diffuom}
  D(\omega)=\left(\begin{array}{cccc}
    0 & 0 & 0 & 0 \\
     0 & \frac{\gamma_m \omega}{\omega_m} \coth\left(\frac{\hbar \omega}{2k_BT}\right) & 0 & 0 \\
    0 & 0 & \kappa & 0 \\
    0 & 0 & 0 & \kappa
  \end{array}\right).
\end{equation}
The $\delta(\omega+\omega')$ factor is a consequence of the stationarity of the noises, which implies the stationarity of the CM $V$: in fact,
inserting Eq.~(\ref{lemma}) into Eq.~(\ref{defV}), one gets that $V$ is time-independent and can be written as
\begin{equation}
V=\int d\omega \tilde{M}(\omega)D(\omega)\tilde{M}(\omega)^{\dagger}. \label{Vfin}
\end{equation}
It is however reasonable to simplify this exact expression for the steady state CM, by appropriately approximating the thermal noise
contribution $D_{22}(\omega)$ in Eq.~(\ref{diffuom}). In fact $k_B T/\hbar \simeq 10^{11}$ s$^{-1}$ even at cryogenic temperatures and it is
therefore much larger than all the other typical frequency scales, which are at most of the order of $10^9$ Hz. The integrand in
Eq.~(\ref{Vfin}) goes rapidly to zero at $\omega \sim 10^{11}$ Hz, and therefore one can safely neglect the frequency dependence of
$D_{22}(\omega)$ by approximating it with its zero-frequency value
\begin{equation}  \label{thermappr}
\frac{\gamma_m \omega}{\omega_m} \coth\left(\frac{\hbar \omega}{2k_BT}%
\right) \simeq \gamma_m \frac{2k_B T}{\hbar \omega_m} \simeq \gamma_m\left(2%
\bar{n}+1\right),
\end{equation}
where $\bar{n}=\left(\exp\{\hbar \omega_m/k_BT\}-1\right)^{-1}$ is the mean thermal excitation number of the resonator.

It is easy to verify that assuming a frequency-independent diffusion matrix $D$ is equivalent to make the following Markovian approximation on
the quantum Brownian noise $\xi(t)$,
\begin{equation}
\label{browncorreMa} \left\langle \xi(t)\xi(t^{\prime})+\xi(t^{\prime})\xi(t)\right\rangle/2 \simeq \gamma_m (2n+1) \delta(t-t^{\prime}),
\end{equation}
which is known to be valid also in the limit of a very high mechanical quality factor ${\cal Q}=\omega_m/\gamma_m \to \infty$ \cite{benguria}.
Within this Markovian approximation, the above frequency domain treatment is equivalent to the time domain derivation considered in \cite{prl07}
which, starting from the formal solution of Eq.~(\ref{compaeq}), arrives at
\begin{equation} \label{cm2}
V_{ij}(\infty)=\sum_{k,l}\int_0^{\infty} ds \int_0^{\infty}ds' M_{ik}(s) M_{jl}(s')D_{kl}(s-s'),
\end{equation}
where $D_{kl}(s-s')=\left(\langle n_k(s)n_l(s')+ n_l(s')n_k(s)\rangle\right)/2$ is the matrix of the stationary noise correlation functions. The
Markovian approximation of the thermal noise on the mechanical resonator yields $D_{kl}(s-s')= D_{kl} \delta(s-s')$, with $D = \mathrm{Diag }
[0,\gamma_m (2 \bar{n}+1),\kappa,\kappa]$, so that Eq.~(\ref{cm2}) becomes
\begin{equation} \label{cm3}
V =\int_0^{\infty} ds  M(s)D M(s)^{T},
\end{equation}
which is equivalent to Eq.~(\ref{Vfin}) whenever $D$ does not depend upon $\omega$. When the stability conditions are satisfied ($M(\infty)=0$),
Eq.~(\ref{cm3}) is equivalent to the following Lyapunov equation for the steady-state CM,
\begin{equation} \label{lyap}
AV+VA^{T}=-D,
\end{equation}
which is a linear equation for $V$ and can be straightforwardly solved, but the general exact expression is too cumbersome and will not be
reported here.

\section{Optomechanical entanglement with the intracavity mode}

\label{Sec:intra}

In order to establish the conditions under which the optical mode and the
mirror vibrational mode are entangled we consider the logarithmic negativity
$E_{\mathcal{N}}$, which can be defined as \cite{logneg}
\begin{equation}
E_{\mathcal{N}}=\max [0,-\ln 2\eta ^{-}],  \label{logneg}
\end{equation}%
where $\eta ^{-}\equiv 2^{-1/2}\left[ \Sigma (V)-\left[ \Sigma (V)^{2}-4\det
V\right] ^{1/2}\right] ^{1/2}$, 
with $\Sigma (V)\equiv \det V_{m}+\det V_{c}-2\det V_{mc}$, and we have used
the $2\times 2$ block form of the CM
\begin{equation}
V\equiv \left(
\begin{array}{cc}
V_{m} & V_{mc} \\
V_{mc}^{T} & V_{c}%
\end{array}%
\right) .  \label{blocks}
\end{equation}%
Therefore, a Gaussian state is entangled if and only if $\eta ^{-}<1/2$,
which is equivalent to Simon's necessary and sufficient entanglement
non-positive partial transpose criterion for Gaussian states \cite{simon},
which can be written as $4\det V<\Sigma -1/4$.

\subsection{Correspondence with the down-conversion process}

As already shown in \cite{marquardt,wilson-rae,genes07} many features of the radiation pressure interaction in the cavity can be understood by
considering that the driving laser light is scattered by the vibrating cavity boundary mostly at the first Stokes ($\omega _{0}-\omega _{m}$)
and anti-Stokes ($\omega _{0}+\omega _{m}$) sidebands. Therefore we expect that the optomechanical interaction and eventually entanglement will
be enhanced
when the cavity is resonant with one of the two sidebands, i.e., when $%
\Delta =\pm \omega _{m}.$

It is useful to introduce the mechanical annihilation operator $\delta
b=(\delta q+i\delta p)/\sqrt{2}$, obeying the following QLE
\begin{equation}
\delta \dot{b}=-i\omega _{m}\delta b-\frac{\gamma _{m}}{2}\left( \delta
b-\delta b^{\dagger }\right) +i\frac{G}{2}\left( \delta a^{\dag }+\delta
a\right) +\frac{\xi }{\sqrt{2}}.  \label{bequat}
\end{equation}
Moving to another interaction picture by introducing the slowly-moving tilded operators $\delta b(t)=\delta \tilde{b}(t)e^{-i\omega _{m}t}$ and
$ \delta a(t)=\delta \tilde{a}(t)e^{-i\Delta t}$, we obtain from the linearized version of Eq.~(\ref{nonlinlang}c) and Eq.~(\ref{bequat}) the
following QLEs
\begin{eqnarray}
\delta \dot{\tilde{b}} &=&-\frac{\gamma _{m}}{2}\left( \delta \tilde{b}%
-\delta \tilde{b}^{\dagger }e^{2i\omega _{m}t}\right) +\sqrt{\gamma _{m}}%
b^{in}  \notag \\
&&\ \ +i\frac{G}{2}\left( \delta \tilde{a}^{\dag }e^{i(\Delta +\omega
_{m})t}+\delta \tilde{a}e^{i(\omega _{m}-\Delta )t}\right)   \label{mode2} \\
\delta \dot{\tilde{a}} &=&-\kappa \delta \tilde{a}+i\frac{G}{2}\left( \delta
\tilde{b}^{\dag }e^{i(\Delta +\omega _{m})t}+\delta \tilde{b}e^{i(\Delta
-\omega _{m})t}\right)   \notag \\
&&\ \ +\sqrt{2\kappa }\tilde{a}^{in}.  \label{modemech2}
\end{eqnarray}%
Note that we have introduced two noise operators: i) $\tilde{a}%
^{in}(t)=a^{in}(t)e^{i\Delta t}$, possessing the same correlation function
as $a^{in}(t)$; ii) $b^{in}(t)=\xi (t)e^{i\omega _{m}t}/\sqrt{2}$ which, in
the limit of large $\omega _{m}$, acquires the correlation functions \cite%
{gard2}
\begin{eqnarray}
\langle b^{in,\dag }(t)b^{in}(t^{\prime })\rangle  &=&\bar{n}\delta
(t-t^{\prime }), \\
\langle b^{in}(t)b^{in,\dag }(t^{\prime })\rangle  &=&\left[ \bar{n}+1\right]
\delta (t-t^{\prime }).
\end{eqnarray}%
Eqs.~(\ref{mode2})-(\ref{modemech2}) are still equivalent to the linearized QLEs of Eq.~(\ref{compaeq}), but now we particularize them by
choosing $\Delta =\pm \omega _{m}.$
If the cavity is resonant with the Stokes sideband of the driving laser, $%
\Delta =-\omega _{m}$, one gets
\begin{eqnarray}
\delta \dot{\tilde{b}} &=&-\frac{\gamma _{m}}{2}\delta \tilde{b}+\frac{%
\gamma _{m}}{2}\delta \tilde{b}^{\dagger }e^{2i\omega _{m}t}+i\frac{G}{2}%
\delta \tilde{a}^{\dag }\nonumber \\
&+& i\frac{G}{2}\delta \tilde{a}e^{2i\omega _{m}t}+%
\sqrt{\gamma _{m}}b^{in},  \label{mode3} \\
\delta \dot{\tilde{a}} &=&-\kappa \delta \tilde{a}+i\frac{G}{2}\delta \tilde{%
b}^{\dag }+i\frac{G}{2}\delta \tilde{b}e^{2i\omega _{m}t}+\sqrt{2\kappa }%
\tilde{a}^{in},  \label{mode4}
\end{eqnarray}%
while when the cavity is resonant with the anti-Stokes sideband of the driving laser, $\Delta =\omega _{m}$, one gets
\begin{eqnarray}
\delta \dot{\tilde{b}} &=&-\frac{\gamma _{m}}{2}\delta \tilde{b}+\frac{%
\gamma _{m}}{2}\delta \tilde{b}^{\dagger }e^{2i\omega _{m}t}+i\frac{G}{2}%
\delta \tilde{a}\nonumber \\
&+& i\frac{G}{2}\delta \tilde{a}^{\dag }e^{-2i\omega _{m}t}+%
\sqrt{\gamma _{m}}b^{in},  \label{mode3prime} \\
\delta \dot{\tilde{a}} &=&-\kappa \delta \tilde{a}+i\frac{G}{2}\delta \tilde{%
b}+i\frac{G}{2}\delta \tilde{b}^{\dag }e^{-2i\omega _{m}t}+\sqrt{2\kappa }%
\tilde{a}^{in}.  \label{mode4prime}
\end{eqnarray}%
From Eqs.~(\ref{mode3})-(\ref{mode4}) we see that, for a blue-detuned driving laser, $\Delta =-\omega _{m}$, the cavity mode and mechanical
resonator are coupled via two kinds of interactions: i) a down-conversion process characterized by $\delta \tilde{b}^{\dag }\delta
\tilde{a}^{\dag }+\delta \tilde{a}\delta \tilde{b}$, which is resonant and ii) a beam-splitter-like process characterized by $\delta
\tilde{b}^{\dagger }\delta \tilde{a}+\delta \tilde{a}^{\dagger }\delta \tilde{b}$, which is off resonant. Since the beam splitter interaction is
not able to entangle modes starting from classical input states \cite{kim}, and it is also off-resonant in this case, one can invoke the
rotating wave approximation (RWA) (which is justified in the limit of $\omega _{m} \gg G,\kappa $) and simplify the interaction to a down
conversion process, which is known to generate bipartite entanglement. In the red-detuned driving laser case,
Eqs.~(\ref{mode3prime})-(\ref{mode4prime}) show that the two modes are strongly coupled by a beam-splitter-like interaction, while the
down-conversion process is off-resonant. If one chose to make the RWA in this case, one would be left with an effective beam splitter
interaction which cannot entangle. Therefore, in the RWA limit $\omega _{m} \gg G,\kappa $, the best regime for strong optomechanical
entanglement is when the laser is blue-detuned from the cavity resonance and down-conversion is enhanced.  However, as it will be seen in the
following section, this is hindered by instability and one is rather forced to work in the opposite regime of a red-detuned laser where
instability takes place only at large values of $G$.

\subsection{Entanglement in the blue-detuned regime}

The CM of the Gaussian steady state of the bipartite system, can be obtained
from Eqs.~(\ref{mode3})-(\ref{mode4}) and Eqs.~Eqs.~(\ref{mode3prime})-(\ref%
{mode4prime}) in the RWA limit, with the techniques of the former section
(see also \cite{gerard})
\begin{equation}
V\equiv V^{\pm }=\left(
\begin{array}{cccc}
V_{11}^{\pm } & 0 & 0 & V_{14}^{\pm } \\
0 & V_{11}^{\pm } & \pm V_{14}^{\pm } & 0 \\
0 & \pm V_{14}^{\pm } & V_{33}^{\pm } & 0 \\
V_{14}^{\pm } & 0 & 0 & V_{33}^{\pm }%
\end{array}%
\right) ,  \label{corremat2}
\end{equation}%
where the upper (lower) sign corresponds to the blue- (red-)detuned case,
and
\begin{subequations}
\label{corrematelemrwa}
\begin{eqnarray}
V_{11}^{\pm } &=&\bar{n}+\frac{1}{2}+\frac{2G^{2}\kappa \left[ 1/2\pm \left(
\bar{n}+1/2\right) \right] }{\left( \gamma _{m}+2\kappa \right) \left(
2\gamma _{m}\kappa \mp G^{2}\right) }, \\
V_{33}^{\pm } &=&\frac{1}{2}+\frac{G^{2}\gamma _{m}\left[ \bar{n}+1/2\pm 1/2%
\right] }{\left( \gamma _{m}+2\kappa \right) \left( 2\gamma _{m}\kappa \mp
G^{2}\right) }, \\
V_{14}^{\pm } &=&\frac{2G\gamma _{m}\kappa \left[ \bar{n}+1/2\pm 1/2\right]
}{\left( \gamma _{m}+2\kappa \right) \left( 2\gamma _{m}\kappa \mp
G^{2}\right) }.
\end{eqnarray}%
For clarity we have included the red-detuned case in the RWA approximation and we see that $\det V_{mc}^{\pm }=\mp (V_{14}^{\pm })^{2}$, i.e.,
is non-negative in this latter case, which is a sufficient condition for the separability of bipartite states \cite{simon}. Of course, this is
expected, since it is just the beam-splitter interaction that generates this CM. Thus, in the weak optomechanical coupling regime of the RWA
limit, entanglement is obtained only for a blue-detuned laser, $\Delta =-\omega _{m} $. However, the amount of achievable optomechanical
entanglement at the steady state is seriously limited by the stability condition of Eq.~(\ref%
{stab}a), which in the RWA limit $\Delta =-\omega _{m}\gg \kappa ,\gamma _{m} $, simplifies to $G<\sqrt{2\kappa \gamma _{m}}$. Since one needs
small mechanical dissipation rate $\gamma _{m}$ in order to see quantum effects, this means a very low maximum value for $G$. The logarithmic
negativity $E_{\mathcal{N}}$ is an increasing function of the effective optomechanical coupling $G$ (as expected) and therefore the stability
condition puts a strong upper bound also on $E_{\mathcal{N}}$. In fact, it is possible to prove that the following bound on $E_{\mathcal{N}}$
exists
\end{subequations}
\begin{equation}
E_{\mathcal{N}}\leq \ln \left[ \frac{1+G/\sqrt{2\kappa \gamma _{m}}}{1+\bar{n%
}}\right] ,  \label{logneg2}
\end{equation}%
showing that $E_{\mathcal{N}}\leq \ln 2$ and above all that entanglement is extremely fragile with respect to temperature in the RWA limit
because, due to the stability condition, $E_{\mathcal{N}}$ vanishes as soon as $\bar{n} \geq 1$.

\subsection{Entanglement in the red-detuned regime}

\begin{figure}[tbh]
\centerline{\includegraphics[width=0.45\textwidth]{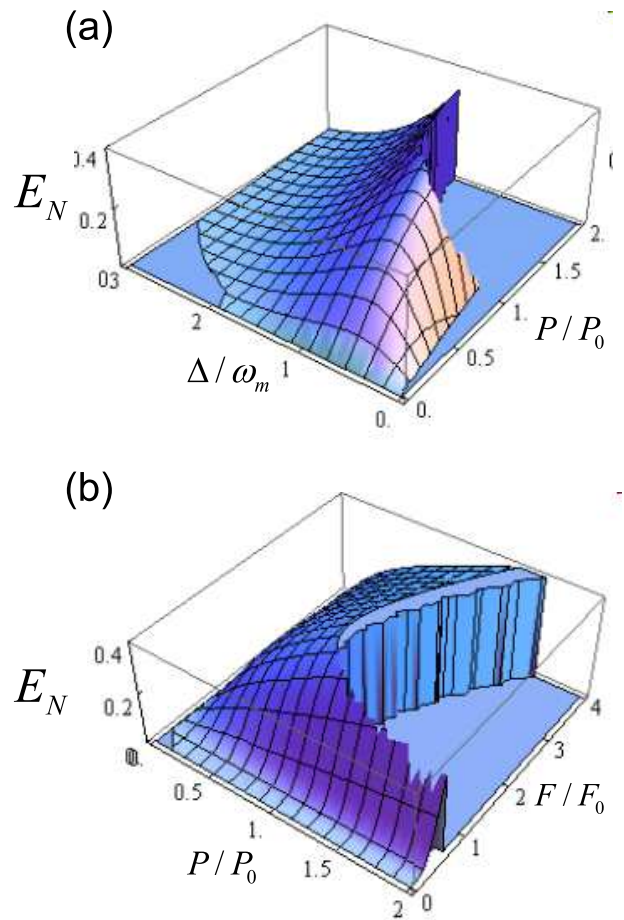}} \caption{(Color online) (a) Logarithmic negativity $E_{\mathcal{N}}$
versus the normalized detuning $\Delta/\protect\omega_m$ and normalized input power $P/P_0$, ($P_0=50$ mW) at a fixed value of the cavity
finesse $F=F_0=1.67 \times 10^4$; (b) $E_{\mathcal{N}}$ versus the normalized finesse $F/F_0$
and normalized input power $P/P_0$ at a fixed detuning $\Delta =\protect%
\omega_m$. Parameter values are $\protect\omega _{m}/2\protect\pi =10$ MHz, $%
\mathcal{Q}=10^5$, mass $m=10$ ng, a cavity of length $L=1$ mm driven by a
laser with wavelength $810$ nm, yielding $G_0=0.95$ KHz and a cavity
bandwidth $\protect\kappa=0.9 \protect\omega_m$ when $F=F_0$. We have
assumed a reservoir temperature for the mirror $T=0.4$ K, corresponding to $%
\bar{n}\simeq 833$. The sudden drop to zero of $E_{\mathcal{N}}$ corresponds
to entering the instability region.}
\label{intracav-ent}
\end{figure}

We conclude that, due to instability, one can find significant
optomechanical entanglement, which is also robust against temperature, only
far from the RWA regime, in the strong coupling regime in the region with
positive $\Delta $, because Eq.~(\ref{stab}b) allows for higher values of $G$%
. This is confirmed by Fig.~\ref{intracav-ent}, where $E_{\mathcal{N}}$ is plotted versus the normalized detuning $\Delta /\omega _{m}$ and the
normalized input power $P/P_{0}$, ($P_{0}=50$ mW) at a fixed value of the cavity finesse $F=F_{0}=1.67\times 10^{4}$ in (a), and versus the
normalized finesse $F/F_{0}$ and normalized input power $P/P_{0}$ at a fixed cavity detuning $\Delta =\omega _{m}$ in (b). We have assumed an
experimentally achievable situation, i.e., a mechanical mode with $\omega _{m}/2\pi =10$ MHz, $\mathcal{Q}=10^{5}$, mass $m=10$ ng, and a cavity
of length $L=1$ mm, driven by a laser with wavelength $810$ nm, yielding $G_{0}=0.95$ kHz and a cavity bandwidth $\kappa =0.9\omega _{m}$ when
$F=F_{0}$. We have assumed a
reservoir temperature for the mirror $T=0.4$ K, corresponding to $\bar{n}%
\simeq 833$. Fig.~\ref{intracav-ent}a shows that $E_{\mathcal{N}}$ is peaked around $\Delta \simeq \omega _{m}$, even though the peak shifts to
larger values of $\Delta $ at larger input powers $P$. For increasing $P$ at fixed $%
\Delta $, $E_{\mathcal{N}}$ increases, even though at the same time the instability region (where the plot suddenly interrupts) widens. In
Fig.~\ref{intracav-ent}b we have fixed the detuning at $\Delta =\omega _{m}$ (i.e., the cavity is resonant with the anti-Stokes sideband of the
laser) and varied
both the input power and the cavity finesse. We see again that $E_{\mathcal{N%
}}$ is maximum just at the instability threshold and also that, once that
the finesse has reached a sufficiently large value, $F\simeq F_{0}$, $E_{%
\mathcal{N}}$ roughly saturates at larger values of $F$. That is, one gets
an optimal optomechanical entanglement when $\kappa \simeq \omega _{m}$ and
moving into the well-resolved sideband limit $\kappa \ll \omega _{m}$ does
not improve the value of $E_{\mathcal{N}}$. The parameter region analyzed is
analogous to that considered in \cite{prl07}, where it has been shown that
this optomechanical entanglement is extremely robust with respect to the
temperature of the reservoir of the mirror, since it persists to more than $%
20$ K.

\subsection{Relationship between entanglement and cooling}

As discussed in detail in \cite{marquardt,wilson-rae,genes07,dantan07} the same cavity-mechanical resonator system can be used for realizing
cavity-mediated optical cooling of the mechanical resonator via the back-action of the cavity mode \cite{brag}. In particular, back-action
cooling is optimized just in the same regime where $\Delta \simeq \omega _{m} $. This fact is easily explained by taking into account the
scattering of the laser light by the oscillating mirror into the Stokes and anti-Stokes sidebands. The generation of an anti-Stokes photon takes
away a vibrational phonon and is responsible for cooling, while the generation of a Stokes photon heats the mirror by producing an extra phonon.
If the cavity is resonant with the anti-Stokes sideband, cooling prevails and one has a positive net laser cooling rate given by the difference
of the scattering rates.

It is therefore interesting to discuss the relation between optimal
optomechanical entanglement and optimal cooling of the mechanical resonator.
This can easily performed because the steady state CM $V$ determines also
the resonator energy, since the effective stationary excitation number of
the resonator is given by $n_{eff}=\left( V_{11}+V_{22}-1\right) /2$ (see
Ref.~\cite{genes07} for the exact expression of these matrix elements giving
the steady state position and momentum resonator variances). In Fig.~\ref%
{intracav-cool} we have plotted $n_{eff}$ under \emph{exactly the same} parameter conditions of Fig.~\ref{intracav-ent}. We see that
\emph{ground state cooling is approached ($n_{eff}<1$) simultaneously with a significant entanglement}. This shows that a significant
back-action cooling of the resonator by the cavity mode is an important condition for achieving an entangled steady state which is robust
against the effects of the resonator thermal bath.

\begin{figure}[tbh]
\centerline{\includegraphics[width=0.45\textwidth]{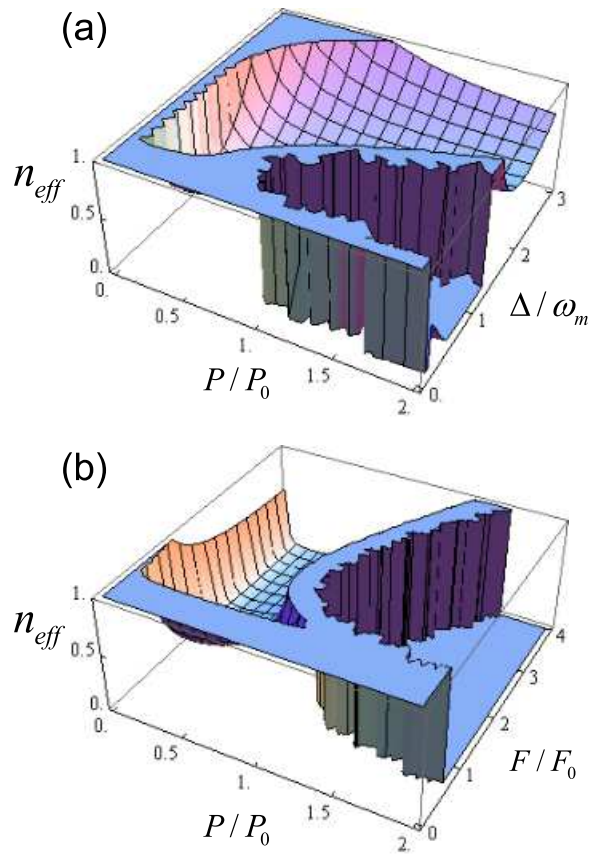}} \caption{(Color online) (a) Effective stationary excitation number of
the
resonator $n_{eff}$ versus the normalized detuning $\Delta /\protect\omega %
_{m}$ and normalized input power $P/P_{0}$, ($P_{0}=50$ mW) at a fixed value
of the cavity finesse $F=F_{0}=1.67\times 10^{4}$; (b) $n_{eff}$ versus the
normalized finesse $F/F_{0}$ and normalized input power $P/P_{0}$ at a fixed
detuning $\Delta =\protect\omega _{m}$. Parameter values are the same as in
Fig.~\protect\ref{intracav-ent}. Again, the sudden drop to zero corresponds
to entering the instability region.}
\label{intracav-cool}
\end{figure}

Nonetheless, entanglement and cooling are different phenomena and optimizing one does not generally optimize also the other. This can be
seen by comparing Figs.~\ref{intracav-ent} and \ref{intracav-cool}: $E_{%
\mathcal{N}}$ is maximized always just at the instability threshold, i.e., for the maximum possible optomechanical coupling, while this is not
true for $n_{eff}$, which is instead minimized quite far from the instability threshold. For a more clear understanding we make use of some of
the results obtained for ground state cooling in Refs.~\cite{marquardt,wilson-rae,genes07}. In the perturbative limit where $G \ll \omega
_{m},\kappa $, one can define scattering rates into the Stokes ($A_{+}$) and anti-Stokes ($A_{-}$) sidebands as
\begin{equation}
A_{\pm }=\frac{G^{2}\kappa /2}{\kappa ^{2}+(\Delta \pm \omega _{m})^{2}},
\end{equation}
so that the net laser cooling rate is given by
\begin{equation}\label{netlaser}
\Gamma =A_{-}-A_{+}>0.
\end{equation}
The final occupancy of the mirror mode is consequently given by \cite{marquardt,wilson-rae,genes07}
\begin{equation}
n_{eff}=\frac{\gamma _{m}\bar{n}}{\gamma _{m}+\Gamma }+\frac{A_{+}}{\gamma
_{m}+\Gamma },
\end{equation}
where the first term in the right hand side of the above equation is the minimized thermal noise, that can be made vanishingly small provided
that $\gamma_m \ll \Gamma$, while the second term shows residual heating produced by Stokes scattering off the vibrational ground state. When
$\Gamma \gg \gamma _{m}\bar{n}$, the lower bound for $n_{eff}$ is practically set by the ratio $A_{+}/\Gamma $. However, as soon as $G$ is
increased for improving the entanglement generation, scattering into higher order sidebands takes place, with rates proportional to higher
powers of $G$. As a consequence, even though the effective thermal noise is still close to zero, residual scattering off the ground state takes
place at a rate that can be much higher than $A_{+}$. This can be seen more clearly in the exact expression of $\langle \delta q^{2}\rangle
=V_{11}$ given in \cite{genes07}, which is shown to diverge at the threshold given by Eq.~(\ref{stab}b).
\section{Optomechanical entanglement with cavity output modes}

The above analysis of the entanglement between the mechanical mode of interest and the intracavity mode provides a detailed description of the
internal dynamics of the system, but it is not of direct use for practical applications. In fact, one typically does not have direct access to
the intracavity field, but one detects and manipulates only the cavity output field. For example, for any quantum communication application, it
is much more important to analyze the entanglement of the mechanical mode with the \emph{optical cavity output}, i.e., how the intracavity
entanglement is transferred to the output field. Moreover, considering the output field provides further options. In fact, by means of spectral
filters, one can always select many different traveling output modes originating from a single intracavity mode and this gives the opportunity
to easily produce and manipulate a multipartite system, eventually possessing multipartite entanglement.

\subsection{General definition of cavity output modes}

The intracavity field $\delta a(t)$ and its output are related by the usual input-output relation \cite{gard}
\begin{equation}  \label{inout}
a^{out}(t)= \sqrt{2\kappa}\delta a(t)-  a^{in}(t),
\end{equation}
where the output field possesses the same correlation functions of the optical input field $a^{in}(t)$ and the same commutation relation, i.e.,
the only nonzero commutator is $\left[a^{out}(t),a^{out}(t ^{\prime})^{\dagger}\right]=\delta(t-t^{\prime})$. From the continuous output field
$a^{out}(t)$ one can extract many independent optical modes, by selecting different time intervals or equivalently, different frequency
intervals (see e.g. \cite{fuchs}). One can define a generic set of $N$ output modes by means of the corresponding annihilation operators
\begin{equation}
a^{out}_k(t)=\int_{-\infty}^{t}ds g_k(t-s)a^{out}(s),\;\;\;k=1,\ldots N,\label{filter1}
\end{equation}
where $g_k(s)$ is the causal filter function defining the $k$-th output mode. These annihilation operators describe $N$ independent optical
modes when $\left[a^{out}_j(t),a^{out}_k(t)^{\dagger}\right]=\delta_{jk}$, which is verified when \begin{equation} \int_{0}^{\infty}ds g_j(s)^*
g_k(s)=\delta_{jk}\label{filter2},
\end{equation}
i.e., the $N$ filter functions $g_k(t)$ form an orthonormal set of square-integrable functions in $[0,\infty)$. The situation can be
equivalently described in the frequency domain: taking the Fourier transform of Eq.~(\ref{filter1}), one has
\begin{equation}
\tilde{a}^{out}_k(\omega)=\int_{-\infty}^{\infty}\frac{dt}{\sqrt{2\pi}} a^{out}_k(t)e^{i\omega
t}=\sqrt{2\pi}\tilde{g}_k(\omega)a^{out}(\omega),\label{filterFT}
\end{equation}
where $\tilde{g}_k(\omega)$ is the Fourier transform of the filter function.
An explicit example of an orthonormal set of filter functions is given by
\begin{equation}
g_k(t)=\frac{\theta(t)-\theta(t-\tau)}{\sqrt{\tau}}e^{-i\Omega_k t} , \label{filterex}
\end{equation}
($\theta$ denotes the Heavyside step function) provided that $\Omega_k$ and $\tau$ satisfy the condition
\begin{equation}\label{interfer}
\Omega_j-\Omega_k=\frac{2\pi}{\tau}p, \;\;\;{\rm integer} \;\;p.
\end{equation} These functions describe a set of independent optical modes, each centered
around the frequency $\Omega_k$ and with time duration $\tau$, i.e., frequency bandwidth $\sim 1/\tau$, since
\begin{equation}
\tilde{g}_k(\omega)=\sqrt{\frac{\tau}{2\pi}}e^{i(\omega-\Omega_k)\tau/2}\frac{\sin\left[(\omega-\Omega_k)\tau/2\right]}{(\omega-\Omega_k)\tau/2}
. \label{filterex2}
\end{equation}
When the central frequencies differ by an integer multiple of $2\pi/\tau$, the corresponding modes are independent due to the destructive
interference of the oscillating parts of the spectrum.

\subsection{Stationary correlation matrix of output modes}

The entanglement between the output modes defined above and the mechanical mode is fully determined by the corresponding $(2N+2)\times (2N+2)$
CM, which is defined by
\begin{equation}
V^{out}_{ij}(t)=\frac{1}{2}\left\langle u^{out}_i(t) u^{out}_j(t)+u^{out}_j(t) u^{out}_i(t)\right\rangle,\label{defVout}
\end{equation}
where
\begin{eqnarray}
&&u^{out}(t)\label{defuout} \\
&&=\left(\delta q(t),\delta p(t),X_1^{out}(t),Y_1^{out}(t),\ldots,X_N^{out}(t),Y_N^{out}(t)\right)^T \nonumber
\end{eqnarray}
is the vector formed by the mechanical position and momentum fluctuations and by the amplitude
($X_k^{out}(t)=\left[a^{out}_k(t)+a^{out}_k(t)^{\dagger}\right]/\sqrt{2}$), and phase
($Y_k^{out}(t)=\left[a^{out}_k(t)-a^{out}_k(t)^{\dagger}\right]/i\sqrt{2})$ quadratures of the $N$ output modes. The vector $u^{out}(t)$
properly describes $N+1$ independent CV bosonic modes, and in particular the mechanical resonator is independent of (i.e., it commutes with) the
$N$ optical output modes because the latter depend upon the output field at former times only ($s<t$). From the definition of $u^{out}(t)$, of
the output modes of Eq.~(\ref{filter1}), and the input-output relation of Eq.~(\ref{inout}) one can write
\begin{eqnarray}
&&u^{out}_i(t)=\int_{-\infty}^t ds T_{ik}(t-s)u^{ext}_k(s)\nonumber \\
&&-\int_{-\infty}^t ds T_{ik}(t-s)n^{ext}_k(s), \label{inoutgen}
\end{eqnarray}
where
\begin{equation}
u^{ext}(t)=\left(\delta q(t),\delta p(t),X(t),Y(t),\ldots,X(t),Y(t)\right)^T\label{defintra}
\end{equation}
is the $2N+2$-dimensional vector obtained by extending the four-dimensional vector $u(t)$ of the preceding section by repeating $N$ times the
components related to the optical cavity mode, and
\begin{equation}\label{defintranois}
n^{ext}(t) =\frac{1}{\sqrt{2\kappa}}\left(0,0,X_{in}(t),Y_{in}(t),\ldots,X_{in}(t),Y_{in}(t)\right)^T
\end{equation}
is the analogous extension of the noise vector $n(t)$ of the former section without however the noise acting on the mechanical mode. In
Eq.~(\ref{inoutgen}) we have also introduced the $(2N+2)\times (2N+2)$ block-matrix consisting of $N+1$ two-dimensional blocks
\begin{widetext}
\begin{equation}\label{transf}
  T(t)=\left(\begin{array}{ccccccc}
    \delta(t) & 0 & 0 & 0 & 0 & 0 & \ldots\\
     0 & \delta(t) & 0 & 0 & 0 & 0 & \ldots\\
    0 & 0 & \sqrt{2\kappa}{\rm Re}g_1(t) & -\sqrt{2\kappa}{\rm Im}g_1(t) & 0 & 0 & \ldots\\
    0 & 0 & \sqrt{2\kappa}{\rm Im}g_1(t) & \sqrt{2\kappa}{\rm Re}g_1(t)& 0 & 0 & \ldots\\
    0 & 0 &  0 & 0 & \sqrt{2\kappa}{\rm Re}g_2(t) & -\sqrt{2\kappa}{\rm Im}g_2(t) &\ldots\\
    0 & 0 &  0 & 0 & \sqrt{2\kappa}{\rm Im}g_2(t) & \sqrt{2\kappa}{\rm Re}g_2(t)&\ldots\\
  \vdots & \vdots & \vdots & \vdots & \vdots & \vdots & \ldots \end{array}\right).
\end{equation}
\end{widetext}
Using Fourier transforms, and the correlation function of the noises, one can derive the following general expression for the stationary output
correlation matrix, which is the counterpart of the $4\times 4$ intracavity relation of Eq.~(\ref{Vfin})
\begin{widetext}
\begin{equation}
V^{out}=\int d\omega
\tilde{T}(\omega)\left[\tilde{M}^{ext}(\omega)+\frac{P_{out}}{2\kappa}\right]D^{ext}(\omega)\left[\tilde{M}^{ext}(\omega)^{\dagger}+\frac{P_{out}}{2\kappa}
\right]\tilde{T}(\omega)^{\dagger}, \label{Vfinout}
\end{equation}
\end{widetext}
where $P_{out}={\rm Diag}[0,0,1,1,\ldots]$ is the projector onto the $2N$-dimensional space associated with the output quadratures, and we have
introduced the extensions corresponding to the matrices $\tilde{M}(\omega)$ and $D(\omega)$ of the former section,
\begin{equation}
\tilde{M}^{ext}(\omega)=\left(i\omega+A^{ext}\right)^{-1},
\end{equation}
with
\begin{equation}\label{aexten}
  A^{ext}=\left(\begin{array}{ccccccc}
    0 & \omega_m & 0 & 0 & 0 & 0 & \ldots\\
     -\omega_m & -\gamma_m & G & 0 & G & 0 & \ldots\\
    0 & 0 & -\kappa & \Delta & 0 & 0 & \ldots\\
    G & 0 & -\Delta & -\kappa & 0 & 0 & \ldots\\
    0 & 0 &  0 & 0 & -\kappa & \Delta &\ldots\\
    G & 0 &  0 & 0 & -\Delta & -\kappa &\ldots\\
  \vdots & \vdots & \vdots & \vdots & \vdots & \vdots & \ldots \end{array}\right).
\end{equation}
and
\begin{equation}\label{dexten}
  D^{ext}(\omega)=\left(\begin{array}{ccccccc}
    0 & 0 & 0 & 0 & 0 & 0 & \ldots\\
     0 & \frac{\gamma_m \omega}{\omega_m} \coth\left(\frac{\hbar \omega}{2k_BT}\right) & 0 & 0 & 0 & 0 & \ldots\\
    0 & 0 & \kappa & 0 & \kappa & 0 & \ldots\\
    0 & 0 & 0 & \kappa & 0 & \kappa & \ldots\\
    0 & 0 & \kappa & 0 & \kappa & 0 & \ldots\\
    0 & 0 & 0 & \kappa & 0 & \kappa & \ldots\\
  \vdots & \vdots & \vdots & \vdots & \vdots & \vdots & \ldots \end{array}\right).
\end{equation}
A deeper understanding of the general expression for $V^{out}$ of Eq.~(\ref{Vfinout}) is obtained by multiplying the terms in the integral: one
gets
\begin{widetext}
\begin{equation}
V^{out}=\int d\omega
\tilde{T}(\omega)\tilde{M}^{ext}(\omega)D^{ext}(\omega)\tilde{M}^{ext}(\omega)^{\dagger}\tilde{T}(\omega)^{\dagger}+\frac{P_{out}}{2}+\frac{1}{2}
\int d\omega \tilde{T}(\omega)\left[\tilde{M}^{ext}(\omega)R_{out}+R_{out}\tilde{M}^{ext}(\omega)^{\dagger}\right]\tilde{T}(\omega)^{\dagger},
\label{Vfinout2}
\end{equation}
\end{widetext}
where $R_{out}=P_{out} D^{ext}(\omega)/\kappa=D^{ext}(\omega)P_{out}/\kappa$ and we have used the fact that
\begin{equation}
\int \frac{d\omega}{4\kappa^2} \tilde{T}(\omega)P_{out}D^{ext}(\omega)P_{out}\tilde{T}(\omega)^{\dagger}=\frac{P_{out}}{2}. \label{Vfinoutlem}
\end{equation}
The first integral term in Eq.~(\ref{Vfinout2}) is the contribution coming from the interaction between the mechanical resonator and the
intracavity field. The second term gives the noise added by the optical input noise to each output mode. The third term gives the contribution
of the correlations between the intracavity mode and the optical input field, which may cancel the destructive effects of the second noise term
and eventually, even increase the optomechanical entanglement with respect to the intracavity case. We shall analyze this fact in the following
section.

\subsection{A single output mode}

\begin{figure}[tbh]
\centerline{\includegraphics[width=0.45\textwidth]{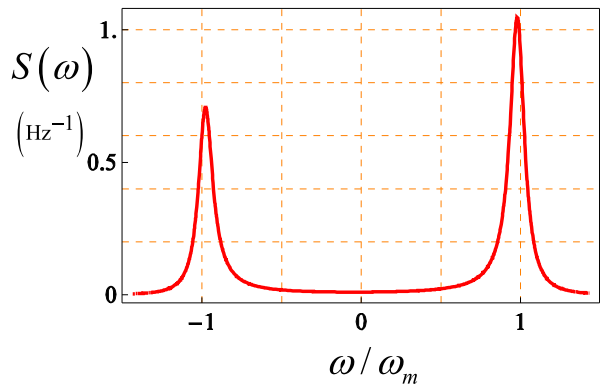}} \caption{(Color online) Cavity output spectrum in the case of an
oscillator with $\omega _{m}/2\pi =10$ MHz, ${\cal Q}=10^5$, mass $m=50$ ng, a cavity of length $L=1$ mm with finesse $F=2 \times 10^4$,
detuning $\Delta =\omega_m$, driven by a laser with input power $P=30$ mW and wavelength $810$ nm, yielding $G_0=0.43$ kHz, $G=0.41 \omega_m$,
and a cavity bandwidth $\kappa=0.75 \omega_m$. We have again assumed a reservoir temperature for the mirror $T=0.4$ K, corresponding to
$\bar{n}\simeq 833$. In this regime photons are scattered only at the two first motional sidebands, at $\omega_0\pm\omega_m$.}
\label{outputspectrum}
\end{figure}

Let us first consider the case when we select and detect only one mode at the cavity output. Just to fix the ideas, we choose the mode specified
by the filter function of Eqs.~(\ref{filterex}) and (\ref{filterex2}), with central frequency $\Omega$ and bandwidth $\tau^{-1}$.
Straightforward choices for this output mode are a mode centered either at the cavity frequency, $\Omega=\omega_c-\omega_0$, or at the driving
laser frequency, $\Omega=0$ (we are in the rotating frame and therefore all frequencies are referred to the laser frequency $\omega_0$), and
with a bandwidth of the order of the cavity bandwidth $\tau^{-1} \simeq \kappa$. However, as discussed above, the motion of the mechanical
resonator generates Stokes and anti-Stokes motional sidebands, consequently modifying the cavity output spectrum. Therefore it may be nontrivial
to determine which is the optimal frequency bandwidth of the output field which carries most of the optomechanical entanglement generated within
the cavity. The cavity output spectrum associated with the photon number fluctuations $S(\omega)= \langle \delta a(\omega)^{\dagger} \delta
a(\omega)\rangle$ is shown in Fig.~\ref{outputspectrum}, where we have considered a parameter regime close to that considered for the
intracavity case, i.e., an oscillator with $\omega _{m}/2\pi =10$ MHz, ${\cal Q}=10^5$, mass $m=50$ ng, a cavity of length $L=1$ mm with finesse
$F=2 \times 10^4$, detuning $\Delta =\omega_m$, driven by a laser with input power $P=30$ mW and wavelength $810$ nm, yielding $G_0=0.43$ kHz,
$G=0.41 \omega_m$, and a cavity bandwidth $\kappa=0.75 \omega_m$. We have again assumed a reservoir temperature for the mirror $T=0.4$ K,
corresponding to $\bar{n}\simeq 833$. This regime is not far but does not corresponds to the best intracavity optomechanical entanglement regime
discussed in Sec.~\ref{Sec:intra}. In fact, optomechanical entanglement monotonically increases with the coupling $G$ and is maximum just at the
bistability threshold, which however is not a convenient operating point. We have chosen instead a smaller input power and a larger mass,
implying a smaller value of $G$ and an operating point not too close to threshold.

\begin{figure}[tbh]
\centerline{\includegraphics[width=0.45\textwidth]{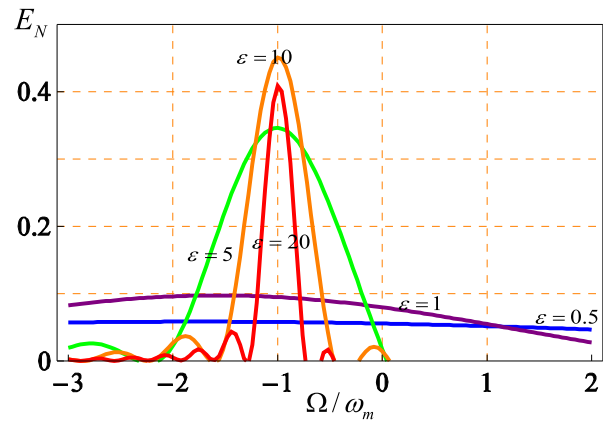}} \caption{(Color online) Logarithmic negativity
$E_{\mathcal{N}}$ of the CV bipartite system formed by mechanical mode and a single cavity output mode versus the central frequency of the
detected output mode $\Omega/\omega_m$ at five different values of its inverse bandwidth $\varepsilon=\omega_m \tau$. The other parameters are
the same as in Fig.~\protect\ref{outputspectrum}. When the bandwidth is not too large, the mechanical mode is significantly entangled only with
the first Stokes sideband at $\omega_0-\omega_m$.} \label{output1}
\end{figure}

In order to determine the output optical mode which is better entangled with the mechanical resonator, we study the logarithmic negativity
$E_{\mathcal{N}}$ associated with the output CM $V^{out}$ of Eq.~(\ref{Vfinout2}) (for $N=1$) as a function of the central frequency of the mode
$\Omega$ and its bandwidth $\tau^{-1}$, at the considered operating point. The results are shown in Fig.~\ref{output1}, where $E_{\mathcal{N}}$
is plotted versus $\Omega/\omega_m$ at five different values of $\varepsilon=\tau \omega_m$. If $\varepsilon \lesssim 1$, i.e., the bandwidth of
the detected mode is larger than $\omega_m$, the detector does not resolve the motional sidebands, and $E_{\mathcal{N}}$ has a value (roughly
equal to that of the intracavity case) which does not essentially depend upon the central frequency. For smaller bandwidths (larger
$\varepsilon$), the sidebands are resolved by the detection and the role of the central frequency becomes important. In particular
$E_{\mathcal{N}}$ becomes highly peaked around the \emph{Stokes sideband} $\Omega=-\omega_m$, showing that the optomechanical entanglement
generated within the cavity is mostly carried by this lower frequency sideband. What is relevant is that the optomechanical entanglement of the
output mode is significantly larger than its intracavity counterpart and achieves its maximum value at the optimal value $\varepsilon \simeq
10$, i.e., a detection bandwidth $\tau^{-1} \simeq \omega_m/10$. This means that in practice, by appropriately filtering the output light, one
realizes an \emph{effective entanglement distillation} because the selected output mode is more entangled with the mechanical resonator than the
intracavity field.

The fact that the output mode which is most entangled with the mechanical resonator is the one centered around the Stokes sideband is also
consistent with the physics of two previous models analyzed in Refs.~\cite{fam,prltelep}. In \cite{fam} an atomic ensemble is inserted within
the Fabry-Perot cavity studied here, and one gets a system showing robust tripartite (atom-mirror-cavity) entanglement at the steady state only
when the atoms are resonant with the Stokes sideband of the laser. In particular, the atomic ensemble and the mechanical resonator become
entangled under this resonance condition, and this is possible only if entanglement is carried by the Stokes sideband because the two parties
are only indirectly coupled through the cavity mode. In \cite{prltelep}, a free-space optomechanical model is discussed, where the entanglement
between a vibrational mode of a perfectly reflecting micro-mirror and the two first motional sidebands of an intense laser beam shined on the
mirror is analyzed. Also in that case, the mechanical mode is entangled only with the Stokes mode and it is not entangled with the anti-Stokes
sideband.

By looking at the output spectrum of Fig.~\ref{outputspectrum}, one can also understand why the output mode optimally entangled with the
mechanical mode has a finite bandwidth $\tau^{-1} \simeq \omega_m/10$ (for the chosen operating point). In fact, the optimal situation is
achieved when the detected output mode overlaps as best as possible with the Stokes peak in the spectrum, and therefore $\tau^{-1}$ coincides
with the width of the Stokes peak. This width is determined by the effective damping rate of the mechanical resonator, $\gamma
_{m}^{eff}=\gamma_m+\Gamma$, given by the sum of the intrinsic damping rate $\gamma_m$ and the net laser cooling rate $\Gamma$ of
Eq.~(\ref{netlaser}. It is possible to check that, with the chosen parameter values, the condition $\varepsilon = 10$ corresponds to
$\tau^{-1}\simeq \gamma _{m}^{eff}$.

\begin{figure}[tbh]
\centerline{\includegraphics[width=0.45\textwidth]{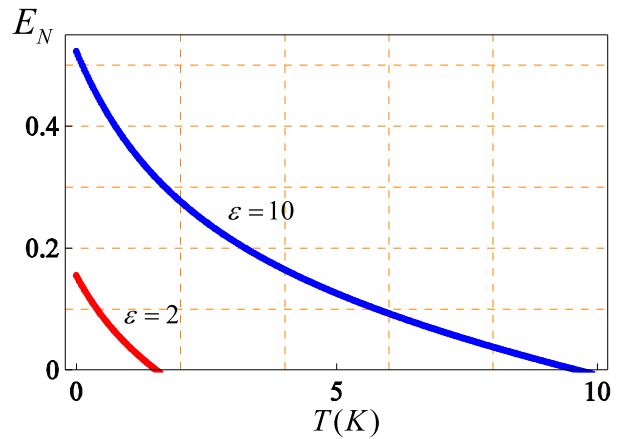}} \caption{(Color online) Logarithmic negativity $E_{\mathcal{N}}$ of the CV
bipartite system formed by mechanical mode and the cavity output mode centered around the Stokes sideband $\Omega=-\omega_m$ versus temperature
for two different values of its inverse bandwidth $\varepsilon=\omega_m \tau$. The other parameters are the same as in
Fig.~\protect\ref{outputspectrum}.} \label{robust-optomech}
\end{figure}

It is finally important to analyze the robustness of the present optomechanical entanglement with respect to temperature. As discussed above and
shown in \cite{prl07}, the entanglement of the resonator with the intracavity mode is very robust. It is important to see if this robustness is
kept also by the optomechanical entanglement of the output mode. This is shown by Fig.~\ref{robust-optomech}, where the entanglement
$E_{\mathcal{N}}$ of the output mode centered at the Stokes sideband $\Omega=-\omega_m$ is plotted versus the temperature of the reservoir at
two different values of the bandwidth, the optimal one $\varepsilon=10$, and at a larger bandwidth $\varepsilon =2$. We see the expected decay
of $E_{\mathcal{N}}$ for increasing temperature, but above all that also this output optomechanical entanglement is robust against temperature
because it persists even above liquid He temperatures, at least in the case of the optimal detection bandwidth $\varepsilon=10$.

\subsection{Two output modes}

Let us now consider the case where we detect at the output two independent, well resolved, optical output modes. We use again the step-like
filter functions of Eqs.~(\ref{filterex}) and (\ref{filterex2}), assuming the same bandwidth $\tau^{-1}$ for both modes and two different
central frequencies, $\Omega_1$ and $\Omega_2$, satisfying the orthogonality condition of Eq.~(\ref{interfer}) $\Omega_1-\Omega_2=2p \pi
\tau^{-1}$ for some integer $p$, in order to have two independent optical modes. It is interesting to analyze the stationary state of the
resulting tripartite CV system formed by the two output modes and the mechanical mode, in order to see if and when it is able to show i) purely
optical bipartite entanglement between the two output modes; ii) fully tripartite optomechanical entanglement.

The generation of two entangled light beams by means of the radiation pressure interaction of these fields with a mechanical element has been
already considered in various configurations. In Ref.~\cite{giovaEPL01}, and more recently in Ref.~\cite{wipf}, two modes of a Fabry-Perot
cavity system with a movable mirror, each driven by an intense laser, are entangled at the output due to their common ponderomotive interaction
with the movable mirror (the scheme has been then generalized to many driven modes in \cite{giannini03}). In the single mirror free-space model
of Ref.~\cite{prltelep}, the two first motional sidebands are also robustly entangled by the radiation pressure interaction as in a two-mode
squeezed state produced by a non-degenerate parametric amplifier \cite{jopbpir}. Robust two-mode squeezing of a bimodal cavity system can be
similarly produced if the movable mirror is replaced by a single ion trapped within the cavity \cite{morigi}.

\begin{figure}[tbh]
\centerline{\includegraphics[width=0.45\textwidth]{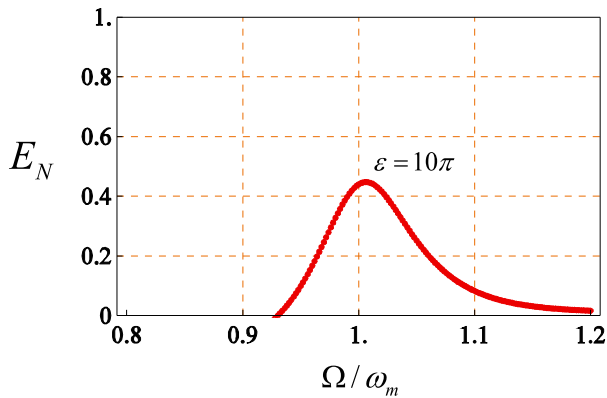}} \caption{(Color online) Logarithmic negativity $E_{\mathcal{N}}$ of the
bipartite system formed by the output mode centered at the Stokes sideband ($\Omega_1=-\omega_m$) and a second output mode with the same inverse
bandwidth ($\varepsilon=\omega_m \tau = 10 \pi$) and a variable central frequency $\Omega$, plotted versus $\Omega/\omega_m$. The other
parameters are the same as in Fig.~\protect\ref{outputspectrum}. Optical entanglement is present only when the second output mode overlaps with
the anti-Stokes sideband.} \label{sideband-ent-sweep}
\end{figure}

\begin{figure}[tbh]
\centerline{\includegraphics[width=0.45\textwidth]{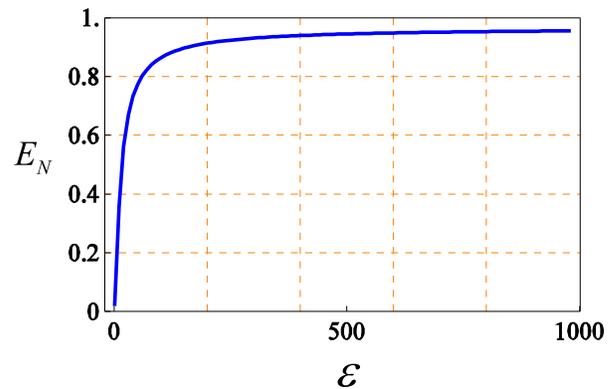}} \caption{(Color online) Logarithmic negativity $E_{\mathcal{N}}$ of the
bipartite system formed by the two output modes centered at the Stokes and anti-Stokes sidebands ($\Omega=\pm\omega_m$) versus the inverse
bandwidth $\varepsilon=\omega_m \tau $. The other parameters are the same as in Fig.~\protect\ref{outputspectrum}.} \label{sideband-ent}
\end{figure}

The situation considered here is significantly different from that of Refs.~\cite{giovaEPL01,wipf,giannini03,morigi}, which require many driven
cavity modes, each associated with the corresponding output mode. In the present case instead, the different output modes \emph{originate from
the same single driven cavity mode}, and therefore it is much simpler from an experimental point of view. The present scheme can be considered
as a sort of ``cavity version'' of the free-space case of Ref.~\cite{prltelep}, where the reflecting mirror is driven by a single intense laser.
Therefore, as in \cite{prltelep,jopbpir}, one expects to find a parameter region where the two output modes centered around the two motional
sidebands of the laser are entangled. This expectation is clearly confirmed by Fig.~\ref{sideband-ent-sweep}, where the logarithmic negativity
$E_{\mathcal{N}}$ associated with the bipartite system formed by the output mode centered at the Stokes sideband ($\Omega_1=-\omega_m$) and a
second output mode with the same inverse bandwidth ($\varepsilon=\omega_m \tau = 10 \pi$) and a variable central frequency $\Omega$, is plotted
versus $\Omega/\omega_m$. $E_{\mathcal{N}}$ is calculated from the CM of Eq.~(\ref{Vfinout2}) (for $N=2$), eliminating the first two rows
associated with the mechanical mode, and assuming the same parameters considered in the former subsection for the single output mode case. One
can clearly see that bipartite entanglement between the two cavity outputs exists only in a narrow frequency interval around the anti-Stokes
sideband, $\Omega=\omega_m$, where $E_{\mathcal{N}}$ achieves its maximum. This shows that, as in \cite{prltelep,jopbpir}, the two cavity output
modes corresponding to the Stokes and anti-Stokes sidebands of the driving laser are significantly entangled by their common interaction with
the mechanical resonator. The advantage of the present cavity scheme with respect to the free-space case of \cite{prltelep,jopbpir} is that the
parameter regime for reaching radiation-pressure mediated optical entanglement is much more promising from an experimental point of view because
it requires less input power and a not too large mechanical quality factor of the resonator. In Fig.~\ref{sideband-ent}, the dependence of
$E_{\mathcal{N}}$ of the two output modes centered at the two sidebands $\Omega=\pm\omega_m$ upon their inverse bandwidth $\varepsilon$ is
studied. We see that, differently from optomechanical entanglement of the former subsection, the logarithmic negativity of the two sidebands
always increases for decreasing bandwidth, and it achieves a significant value ($\sim 1$), comparable to that achievable with parametric
oscillators, for very narrow bandwidths. This fact can be understood from the fact that quantum correlations between the two sidebands are
established by the coherent scattering of the cavity photons by the oscillator, and that the quantum coherence between the two scattering
processes is maximal for output photons with frequencies $\omega_0 \pm \omega_m$.

\begin{figure}[tbh]
\centerline{\includegraphics[width=0.45\textwidth]{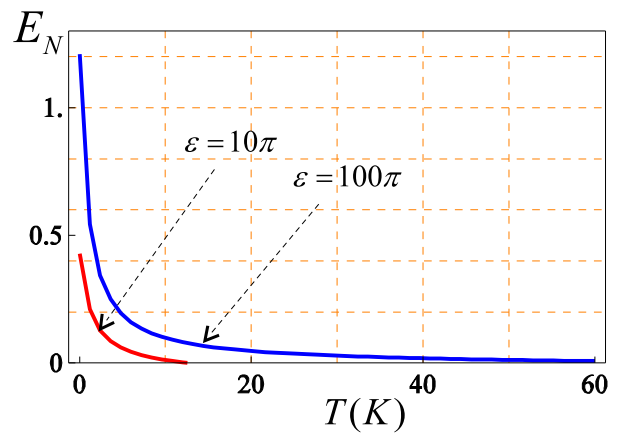}} \caption{(Color online) Logarithmic negativity $E_{\mathcal{N}}$ of
the two output modes centered at the Stokes and anti-Stokes sidebands ($\Omega=\pm\omega_m$) versus the temperature of the resonator reservoir,
at two different values of the inverse bandwidth $\varepsilon=\omega_m \tau $. The other parameters are the same as in
Fig.~\protect\ref{outputspectrum}.} \label{robu-sideband}
\end{figure}

In Fig.~\ref{robu-sideband} we analyze the robustness of the entanglement between the Stokes and anti-Stokes sidebands with respect to the
temperature of the mechanical resonator, by plotting, for the same parameter regime of Fig.~\ref{sideband-ent}, $E_{\mathcal{N}}$ versus the
temperature $T$ at two different values of the inverse bandwidth ($\varepsilon=10\pi, 100\pi$). We see that this purely optical CV entanglement
is extremely robust against temperature, especially in the limit of small detection bandwidth, showing that the effective coupling provided by
radiation pressure can be strong enough to render optomechanical devices with high-quality resonator a possible alternative to parametric
oscillators for the generation of entangled light beams for CV quantum communication.

\begin{figure}[tbh]
\centerline{\includegraphics[width=0.45\textwidth]{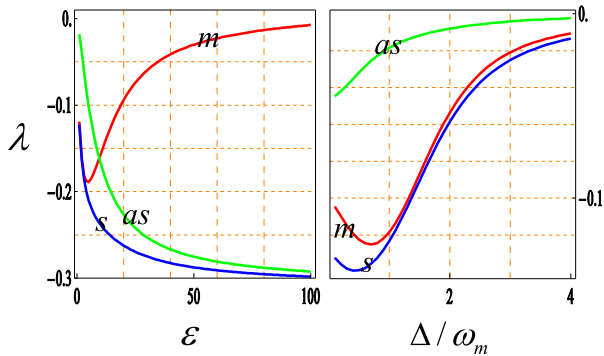}} \caption{(Color online) Analysis of tripartite entanglement. The minimum
eigenvalues after partial transposition with respect to the Stokes mode (blue line), anti-Stokes mode (green line) and the mechanical mode (red
line) are plotted versus the inverse bandwidth $\varepsilon$ at $\Delta=\omega_m$ in the left plot, and versus the cavity detuning
$\Delta/\omega_m$ at the fixed inverse bandwidth $\varepsilon=\pi$ in the right plot. The other parameters are the same as in
Fig.~\protect\ref{outputspectrum}. These eigenvalues are all negative in the studied intervals, showing that one has fully
tripartite-entanglement.} \label{tripartite}
\end{figure}

Since in Figs.~\ref{sideband-ent-sweep} and \ref{sideband-ent} we have used the same parameter values for the cavity-resonator system used in
Fig.~\ref{output1}, we have that in this parameter regime, the output mode centered around the Stokes sideband mode shows bipartite entanglement
simultaneously with the mechanical mode and with the anti-Stokes sideband mode. This fact suggests that, in this parameter region, the CV
tripartite system formed by the output Stokes and anti-Stokes sidebands and the mechanical resonator mode could be characterized by a fully
tripartite-entangled stationary state. This is confirmed by Fig.~\ref{tripartite}, where we have applied the classification criterion of
Ref.~\cite{giedke}, providing a necessary and sufficient criterion for the determination of the entanglement class in the case of tripartite CV
Gaussian states, which is directly computable in terms of the eigenvalues of appropriate test matrices \cite{giedke}. These eigenvalues are
plotted in Fig.~\ref{tripartite} versus the inverse bandwidth $\varepsilon$ at $\Delta=\omega_m$ in the left plot, and versus the cavity
detuning $\Delta/\omega_m$ at the fixed inverse bandwidth $\varepsilon=\pi$ in the right plot (the other parameters are again those of
Fig.~\ref{outputspectrum}). We see that all the eigenvalues are negative in a wide interval of detunings and detection bandwidth of the output
modes, showing, as expected, that we have a fully tripartite-entangled steady state.

Therefore, if we consider the system formed by the two cavity output fields centered around the two motional sidebands at $\omega_0\pm \omega_m$
and the mechanical resonator, we find that the entanglement properties of its steady state are identical to those of the analogous tripartite
optomechanical free-space system of Ref.~\cite{prltelep}. In fact, the Stokes output mode shows bipartite entanglement both with the mechanical
mode and with the anti-Stokes mode, the anti-Stokes mode is not entangled with the mechanical mode, but the whole system is in a fully
tripartite-entangled state for a wide parameter regime. What is important is that in the present cavity scheme, such a parameter regime is much
easier to achieve with respect to that of the free-space case.

\section{Conclusions}

We have studied in detail the entanglement properties of the steady state of a driven optical cavity coupled by radiation pressure to a
micromechanical oscillator, extending in various directions the results of Ref.~\cite{prl07}. We have first analyzed the intracavity steady
state and shown that the cavity mode and the mechanical element can be entangled in a robust way against temperature. We have also investigated
the relationship between entanglement and cooling of the resonator by the back-action of the cavity mode, which has been already demonstrated
recently in Refs.~\cite{gigan06,arcizet06b,vahalacool,mavalvala,sidebcooling,harris,markusepl} and discussed theoretically in
Refs.~\cite{brag,marquardt,wilson-rae,genes07,dantan07}. We have seen that a significant back-action cooling is a sufficient but not necessary
condition for achieving entanglement. In fact, intracavity entanglement is possible also in the opposite regime of negative detunings $\Delta$
where the cavity mode \emph{drives} and does not cool the resonator, even though it is not robust against temperature in this latter case.
Moreover, entanglement is not optimal when cooling is optimal, because the logarithmic negativity is maximized close to the stability threshold
of the system, where instead cooling is not achieved.

We have then extended our analysis to the cavity output, which is more important from a practical point of view because any quantum
communication application involves the manipulation of traveling optical fields. We have developed a general theory showing how it is possible
to define and evaluate the entanglement properties of the multipartite system formed by the mechanical resonator and $N$ independent output
modes of the cavity field.

We have then applied this theory and have seen that in the parameter regime corresponding to a significant intracavity entanglement, the
tripartite system formed by the mechanical element and the two output modes centered at the first Stokes and anti-Stokes sideband of the driving
laser (where the cavity output noise spectrum is concentrated) shows robust fully tripartite entanglement. In particular, the Stokes output mode
is strongly entangled with the mechanical mode and shows a sort of entanglement distillation because its logarithmic negativity is significantly
larger than the intracavity one when its bandwidth is appropriately chosen.

In the same parameter regime, the Stokes and anti-Stokes sideband modes are robustly entangled, and the achievable entanglement in the limit of
a very narrow detection bandwidth is comparable to that generated by a parametric oscillators. These results make the present cavity
optomechanical system very promising for the realization of CV quantum information interfaces and networks.

\section{Acknowledgements}

This work has been supported by the European Commission (programs QAP), and by INFN (SQUALO project).

\end{document}